\documentclass[aps,prb,twocolumn]{revtex4-2}
\usepackage{graphicx}
\usepackage{placeins}
\usepackage{blindtext}
\usepackage[backref=false]{hyperref} 
\usepackage{enumitem}
\usepackage{amsfonts}
\usepackage{amsmath,amsthm}
\usepackage{multirow}
\usepackage{makecell}
\usepackage{diagbox}
\usepackage{xcolor}
\usepackage[braket, qm]{qcircuit}
\usepackage[]{bigints}
\usepackage{enumitem} 
\usepackage{float}

\setlist[enumerate]{leftmargin=*, labelsep=1em, itemsep=0em, topsep=0em, parsep=0pt}

\makeatletter
\renewcommand{\section}{\@startsection{section}{1}{0mm} 
{1.0\baselineskip}{0.01\baselineskip}{\large\bfseries\raggedright}} 
\renewcommand{\subsection}{\@startsection{subsection}{2}{0mm} 
{0.5\baselineskip}{0.01\baselineskip}{\large\bfseries\raggedright}} 
\renewcommand{\subsubsection}{\@startsection{subsection}{2}{0mm} 
{0.5\baselineskip}{0.01\baselineskip}{\normalsize\bfseries\raggedright}} 
\makeatother

\hypersetup{
    colorlinks=true, 
    linkcolor=blue,   
    anchorcolor=blue, 
    citecolor=blue,   
    urlcolor=blue,    
    filecolor=blue    
}

\begin{document}

\title{The inherent convolution property of quantum neural networks}
\author{Guangkai Qu}
\affiliation{
  \begin{minipage}{0.8\textwidth}
    Faculty of Information Science and Engineering, Ocean University of China, Qingdao 266100, China
  \end{minipage}
}

\author{Zhimin Wang}
\thanks{Co-first authors: Guangkai Qu and Zhimin Wang; \\
Corresponding author: wangzhimin@ouc.edu.cn}

\affiliation{
  \begin{minipage}{0.8\textwidth}
    Faculty of Information Science and Engineering, Ocean University of China, Qingdao 266100, China
  \end{minipage}
}

\author{Guoqiang Zhong}
\affiliation{
  \begin{minipage}{0.8\textwidth}
    Faculty of Information Science and Engineering, Ocean University of China, Qingdao 266100, China
  \end{minipage}
}

\author{Yongjian Gu}
\thanks{Corresponding author: yjgu@ouc.edu.cn}
\affiliation{
  \begin{minipage}{0.8\textwidth}
    Faculty of Information Science and Engineering, Ocean University of China, Qingdao 266100, China
  \end{minipage}
}

\begin{abstract}
\noindent Quantum neural networks (QNNs) represent a pioneering intersection of quantum computing and deep learning. In this study, we unveil a fundamental convolution property inherent to QNNs, stemming from the natural parallelism of quantum gate operations on quantum states. Notably, QNNs are capable of performing a convolutional layer using a single quantum gate, whereas classical methods require $2^n$ basic operations. This essential property has been largely overlooked in the design of existing quantum convolutional neural networks (QCNNs), limiting their ability to capture key structural features of classical CNNs, including local connectivity, parameter sharing, and multi-channel, multi-layer architectures. To address these limitations, we propose novel QCNN architectures that explicitly harness the convolutional nature of QNNs. We validate the effectiveness of these architectures through extensive numerical experiments focused on multiclass image classification. Our findings provide deep insights into the realization of convolutional mechanisms within QNNs, marking a substantial advancement in the development of QCNNs and broadening their potential for efficient data processing.
\end{abstract}
\maketitle

\noindent Variational quantum algorithms have emerged as a promising approach for achieving near-term quantum advantages on noisy intermediate-scale quantum (NISQ) devices\textsuperscript{\cite{ref1,ref2,ref3}}. Their conceptual resemblance to artificial neural networks\textemdash involving the optimization of variational ansatzes to solve specific problems\textemdash has led to a popular framework known as quantum neural networks (QNNs)\textsuperscript{\cite{ref4,ref5,ref5-1,ref6,ref7,ref8}}. To date, significant efforts have been made to investigate the trainability and generalization properties of QNNs\textsuperscript{\cite{ref9,ref10,ref11,ref12,ref12-1}}, as well as their applications across diverse domains\textsuperscript{\cite{ref2,ref13}}.

Drawing inspiration from the successes of deep learning models\textsuperscript{\cite{ref14}}, various specialized QNN architectures have been proposed, including deep QNNs\textsuperscript{\cite{ref7,ref8}}, convoultional QNNs\textsuperscript{\cite{ref15,ref16,ref17,ref17-1,ref17-2,ref17-3}}, recurrent QNNs\textsuperscript{\cite{ref18,ref19,ref20,ref21}}, and adversarial generative QNNs \textsuperscript{\cite{ref22,ref23,ref24}}. According to representation learning theroy\textsuperscript{\cite{ref25}}, deep models learn distributed representations of data at multiple levels of abstraction, which helps mitigate issues such as vanishing and exploding gradients. Consequently, QNNs with specialized architectures are expected to address fundamental challenges such as barren plateaus\textsuperscript{\cite{ref25-1,ref25-2,ref25-3}}, and improve generalization performance in domain-specific learning tasks\textsuperscript{\cite{ref17,ref21,ref24,ref26,ref27}}. 

A prominent example is the quantum convolutional neural networks (QCNNs) initially introduced by Cong et al.\textsuperscript{\cite{ref15}}, which seeks to extend the principles of classical CNNs to the quantum domain. These QCNNs are structured with quantum convolutional and pooling layers: the convolutional layers consist of parallel, uniformly parametrized local unitary quantum gates, while the pooling layers employ controlled gates in conjunction with mid-circuit measurements. 
Building upon these QCNNs, hybrid quantum-classical CNNs (also refered to as quanvolutional neural networks) have been introduced\textsuperscript{\cite{ref16,ref33,ref33-1,ref34,ref29-1}}, in which parameterized quantum circuits (PQCs) and classical feedforward neural networks are applied alternately to perform the feedforwad transformation. Notably, QCNNs have been demonstrated to mitigate the barren plateau problem\textsuperscript{\cite{ref26}}, and have shown significant promise in tasks such as quantum phase recognition\textsuperscript{\cite{ref17,ref28,ref28-1}}. 

However, recent studies suggest that QCNNs may not perform as effectively as their classical CNN counterparts in image processing tasks, particularly in multiclass classification scenarios\textsuperscript{\cite{ref29,ref30,ref31,ref32}}. Instead, the hybrid quantum-classical CNN models may offer certain advantages over classical models\textsuperscript{\cite{ref16,ref33,ref33-1,ref34}}. This is particularly intriguing given that classical CNNs are renowned for their proficiency in processing image data, whereas QCNNs seem to fall short in this domain. This discrepancy indicates that QCNNs may fail to fully capture the critical structural features of classical CNNs, such as local connectivity and parameter sharing (which enable local receptive fields), as well as multi-channel and multi-layer architectures (which support distributed and hierarchical feature representations)\textemdash all of which are essential for effective image processing.

In this paper, we show that the operation of quantum gates on quantum states naturally enables QNNs to perform convolutional operations on input data. Specifically, a single parameterized two-qubit gate in QNNs actually performs a convolutional layer with four kernels. However, the existing QCNN designs tend to obscure these intrinsic convolutional operations, limiting their ability to capture the essential structural features of classical CNNs. This limitation contributes to the relative inefficiency of the current QCNNs in image processing tasks. Our findings highlight the importance of evaluating whether the quantum state evolutions in specialized QNN architectures fully exploit the underlying mechanisms of their classical counterparts to achieve superior performance in specific learning tasks. 

\subsection*{Results}
\subsubsection*{Single convolution layer}
\noindent In classical CNNs, the convolution operation is typically simplified to a cross-correlation operation. Specifically, given a filter $W \in \mathbb{R}^{U \times V}$ (oftern referred to as a kernel) that slides across an image $X \in \mathbb{R}^{M \times N}$, where $U(V) \ll M(N)$, the convolution output is generally expressed as:
\begin{equation}\label{Eq0-1}
    y_{ij} = \sum_{u=1}^{U} \sum_{v=1}^{V} w_{uv}x_{i-u+1,j-v+1},
\end{equation}
where $x_{i-u+1,j-v+1}$ represents the pixel in a local image patch $X_i$ (i.e., the receptive field of the kernal), and the indices $(i,j)$ begin from $(U,V)$ for convenience. By flattening the kernal matrix $W$ and the local image patch $X_i$ into row and column vecters, respectively, namely $\boldsymbol{w} \in \mathbb{R}^{1 \times UV}$ and $\boldsymbol{x_i} \in \mathbb{R}^{UV \times 1}$, each element of the convolution output can be written as $y_i = \boldsymbol{w} \cdot \boldsymbol{x_i}$. Furthermore, by constructing a direct sum space, specifically $\bigoplus_{i=1}^{N} \boldsymbol{w} = diag(\boldsymbol{w} , \boldsymbol{w}, \cdots, \boldsymbol{w}) \in \mathbb{R}^{N \times NUV} $, the convolution output can be expressed as:
\begin{equation}\label{Eq0-2}
     \bigoplus_{i=1}^{N}y_i = \left( \bigoplus_{i=1}^{N} \boldsymbol{w} \right) \left( \bigoplus_{i=1}^{N} \boldsymbol{x_i} \right) = \left( I \otimes \boldsymbol{w} \right) \left( \bigoplus_{i=1}^{N} \boldsymbol{x_i}\right) ,
\end{equation}
where $\bigoplus_{i=1}^{N} \boldsymbol{x_i} \in \mathbb{R}^{NUV \times 1}$, and $I \in \mathbb{R}^{N \times N} $ is the identity matrix.

Notably, the transformation in Eq. \ref{Eq0-2} exhibits a natural correspondence to the action of quantum gates on the amplitudes of an input quantum state, making it feasible to implement efficiently via a quantum circuit. Specifically, by combining a set of orthonormal kernels $\{\boldsymbol{w_i}\}$ to construct a unitary transformation $U \in \mathbb{R}^{UV \times UV}$ and setting $N = 2^n$, the term $I \otimes U$ can be efficiently executed through a quantum gate with adjustable parameters. Additionally, the term $\bigoplus_{i=1}^{N} \boldsymbol{x_i}$ corresponds to encoding the image data into the initial quantum state.

\begin{figure*}
    \centering
    \includegraphics[width=1.0\linewidth]{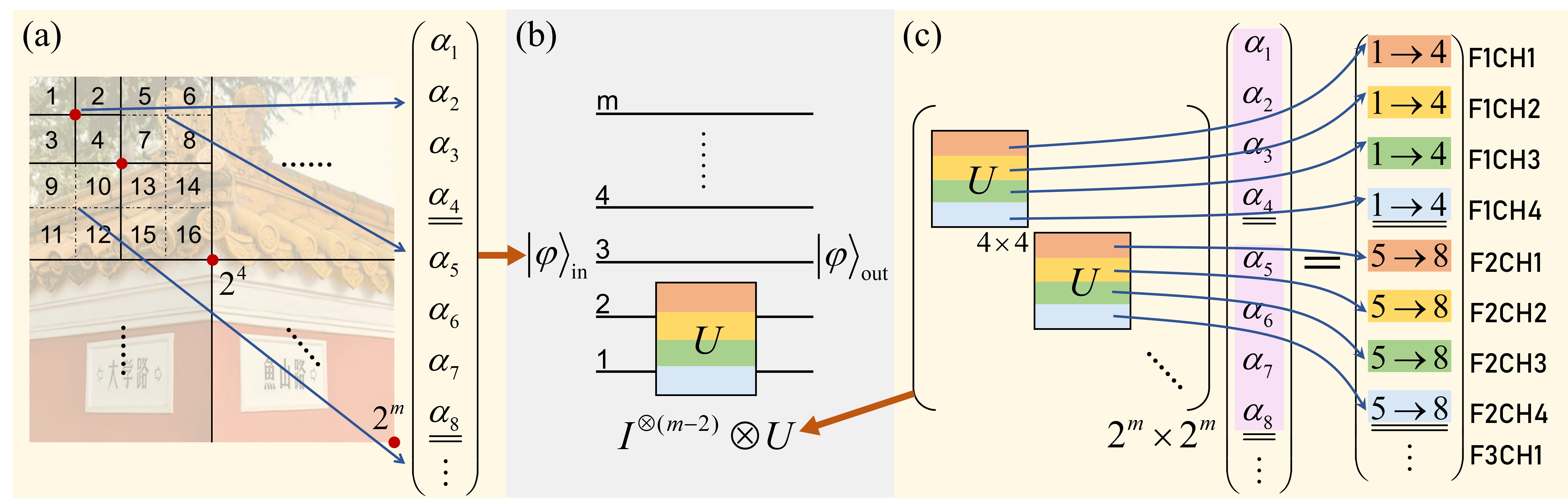}
    \caption{Illustration of the equivalence between a single quantum gate and a convolutional layer. (a) The image pixels are rearranged into a column vector, with each $2 \times 2$ local patch treated as a unit. Specifically, the pixels $1, 2, 3, 4$ in the first local patch are encoded as $\alpha_1, \alpha_2, \alpha_3, \alpha_4$, and the pixels $5, 6, 7, 8$ are encoded as $\alpha_5, \alpha_6, \alpha_7, \alpha_8$, and so on. The size of each local patch corresponds to the receptive field of the convolution layer. This column vector is then encoded into a quantum state using amplitude encoding, resulting in the initial quantum state $\left| \psi \right\rangle_{in}$. (b) A two-qubit quantum gate, denoted as $U$, is applied to the least significant qubits. The four orthonormal row vectors of the $U$ matrix (depicted in four different colors in the figure) serve as the four convolutional kernels, corresponding to the four channels of a classical CNN. Note that the matrix $U$ is derived by expanding $\boldsymbol{w}$ into multiple kernels, as discussed below Eq. \ref{Eq0-2}. (c) The matrix transformation of the quantum circuit in (b) is expressed as a block matrix, where $2^{m-2}$ copies of the $U$-matrix arranged along the diagonal. Each block processes the pixels of a local patch (i.e., the $2 \times 2$ unit). Specifically, the first block processes the first four elements, $\alpha_1$ to $\alpha_4$, and generates the first feature, which is equivalent to performing a filtering operation on the first local region (from pixels $1$ to $4$). Each block contains four channels, with the resulting features denoted as F1CH1, F1CH2, F1CH3 and F1CH4 in the figure (depicted in four different colors corresponding to those of matrix $U$). Similarly, the second block processes the next set of pixels (from $\alpha_5$ to $\alpha_8$) and produces the second feature, represented as F2CH1, F2CH2, F2CH3 and F2CH4 across the four channels. Therefore, a single two-qubit quantum gate performs a convolutional layer with a kernel size of 4, a stride of 4, and 4 output channels.}
    \label{Fig1}
\end{figure*}

A more concrete and graphical illustration of this principle is presented in Fig. \ref{Fig1}. As shown in the figure, a single two-qubit quantum gate within a quantum circuit can be interpreted as equivalent to a convolutional layer with a kernel size of 4, a stride of 4, and 4 output channels (corresponding to the application of 4 distinct kernels). Remarkably, while classical CNNs require $2^n$ basic operations to perform this computation, the quantum model achieves an equivalent outcome with just a single quantum operation.

It is worth noting that each row vector of the transformation matrix corresponding to the quantum gate functions as a convolution kernel. These kernels are intrinsically orthogonally normalized, a direct consequence of the unitary nature of quantum gate evolution\textemdash an attribute absent in classical CNNs. This normalization property ensures the stability of the feedforward process by preventing output values from diverging or diminishing excessively as the number of layers increases. Furthermore, the orthogonality of the kernels promotes a more dispersed distribution of extracted features within the feature space, thereby enhancing their separability. This dispersion contributes to improved quality in the feature representations, which is critical for downstream tasks such as classification and regression.

Fig. \ref{Fig1} illustrates the fundamental scenario where a two-qubit quantum gate operates on the least significant qubits. By adjusting the number and position of active qubits, various convolution operations can be realized. 

First, an $n$-qubit quantum gate, which is actually implemented through a PQC with $n$ qubits, corresponds to a convolutional layer with a kernel size of $2^n$, a stride of $2^n$, and $2^n$ output channels. For an input state comprising $m$ qubits, each output channel contains $2^{m-n}$ features. As the number of qubits of the quantum gate increases, the receptive field of the corresponding convolution operation expands proportionally. Notably, when $m = n$, the kernel spans the entire input, effectively performing a global convolution. In other words, when the ansatz circuit of a QNN is itself a PQC, the QNN can be interpreted as a convolutional network with a global receptive field.

Second, altering the position of the qubits on which the quantum gate acts modifies the size of the receptive field of the corresponding convolution layer. Interestingly, this case actually corresponds to the dilated convolution mechanism (also referred to as atrous convolution) used in classical CNNs\textsuperscript{\cite{ref34-1,ref34-2}}. Specifically, moving the quantum gate upwards from the bottom can be expressed as $U \otimes I_{N_1}$, and then the resulting changes for each convolution kernel $\boldsymbol{w}$ can be descriped as follows:
\begin{equation}\label{Eq0-3}
  \begin{aligned}
     & \left( I_{N_2} \otimes (U \otimes I_{N_1}) \right) \left( \bigoplus_{i=1}^{N} \boldsymbol{x_i}\right) \xrightarrow[\text{kernel}]{\text{one}} \left( \bigoplus_{i=1}^{N_2} \boldsymbol{w^{\prime}} \right) \left( \bigoplus_{i=1}^{N} \boldsymbol{x_i} \right), \\
     & \text{where } \boldsymbol{w^{\prime}} = \boldsymbol{w} \otimes  \boldsymbol{e}. 
  \end{aligned}
\end{equation}
Here, the subscripts $N_1$ and $N_2$ denote the dimensions of the identity matrices, and $\boldsymbol{e} \in \mathbb{R}^{1 \times N_1}$ is the one-hot vector, with one element being 1 and the rest being 0. 

For instance, when a two-qubit quantum gate acts on the two least-significant qubits, as shown in Fig. \ref{Fig1}, $I_{N_1}$ and $\boldsymbol{e_i}$ reduce to scalars of $1$, meaning that each output feature is a weighted summation of the four nearest-neighbor input data points. When the gate is shifted upward by one position to act on the 2nd and 3rd qubits, i.e., when $N_1 = 2$, $\boldsymbol{e} = (1,0)$ (or $\boldsymbol{e} = (0,1)$), and $\boldsymbol{w^{\prime}} = (w_1,0,w_2,0,w_3,0,\cdots)$ (or $\boldsymbol{w^{\prime}} = (0,w_1,0,w_2,0,w_3,\cdots)$), each output feature is now a weighted summation of the four next-nearest-neighbor data points. More generally, shifting the gate upward by $i$ positions corresponds to $N_1 = 2^i$, meaning that each output feature becomes the weighted summation of the four $(2^i)$th-nearest-neighbor data points. This behavior mirrors a dilated convolution layer with a dilation rate of $2^i$.

In general, the relationship between a specific output feature in a channel and the input data points can be formulated as follows:

\textbf{Theorem 1.} For an $n$-qubit quantum gate acting on the $i$th to $(i+n-1)$th qubits of an $m$-qubit input state ($i \in [1,m]$), the operation is equivalent to performing a convolutional layer with a kernel size of $2^n$, a stride of $2^n$, and $2^n$ output channels. The $j$th output feature in channel $c$, denoted as $F_{j,c}$, is given by: 
\begin{equation}\label{Eq1}
   \begin{aligned}
     & F_{j,c} = \sum_{l=1}^{2^n} u_{c,l} \alpha_s, \\
     & \text{where } s = (2^n-1) \cdot 2^{i-1} \cdot \left\lfloor \frac{j-1}{2^{i-1}} \right\rfloor + j + (l-1) \cdot 2^{i-1},\\
     & \text{and } j=1,2,...,2^{m-n}, c=1,2,...2^n.
   \end{aligned}
\end{equation}
Here, $u_{c,l}$ and $\alpha_s$ denote the elements of the quantum gate matrix and the input state vector, respectively. 

\subsubsection*{Multi-convolution layers}
\noindent Having demonstrated that a single quantum gate is equivalent to a single convolutional layer, we now show that two quantum gates can replicate the effect of two consecutive convolutional layers. The key insight lies in the fact that different convolutional layers in classical CNNs possess distinct receptive fields, allowing them to extract features at various levels of abstraction. This hierarchical feature extraction can be emulated by applying quantum gates at different positions within a QNN. Specifically, based on Eq. \ref{Eq0-3}, two parallel quantum gates, denoted as $U_1$ and $U_2$, can be used to construct a composite convolution kernel as follows:
\begin{equation}\label{Eq2-1}
  \begin{aligned}
     & \left( I_{N_2} \otimes (U_2 \otimes U_1) \right) \left( \bigoplus_{i=1}^{N} \boldsymbol{x_i}\right) \xrightarrow[\text{kernel}]{\text{one}} \left( \bigoplus_{i=1}^{N_2} \boldsymbol{w^{\prime}} \right) \left( \bigoplus_{i=1}^{N} \boldsymbol{x_i} \right), \\
     & \text{where } \boldsymbol{w^{\prime}} = \boldsymbol{w_2} \otimes  \boldsymbol{w_1}. 
  \end{aligned}
\end{equation}
Here, $\boldsymbol{w_2}$ and $\boldsymbol{w_1}$ are row vectors of the unitary matrices $U_1$ and $U_2$, respectively, and represent the kernels of the two convolutional layers. Their tensor product, $\boldsymbol{w^{\prime}}$, forms a combined kernel that captures the hierarchical feature transformation achieved by two consecutive convolutional layers. 

\begin{figure*}
    \centering
    \includegraphics[width=1.0\linewidth]{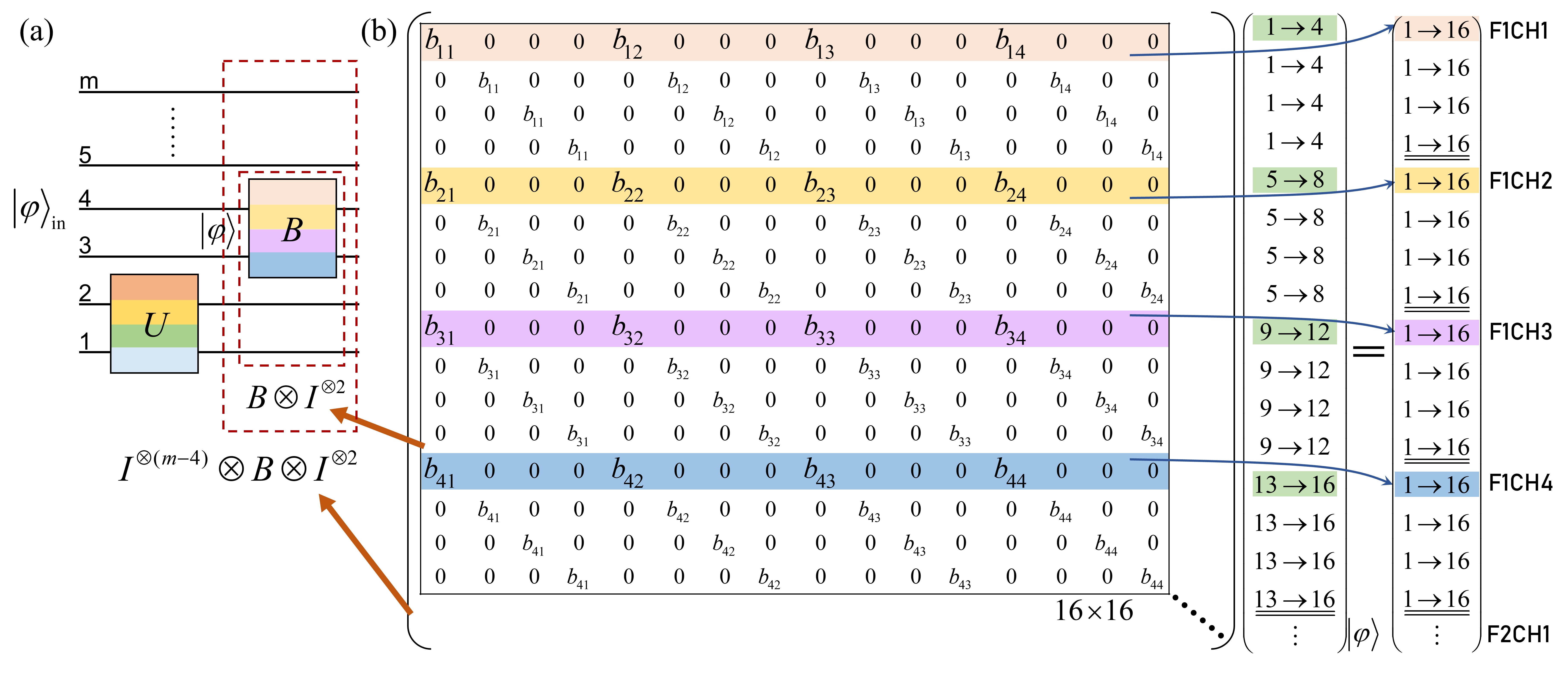}
    \caption{Illustration of the equivalence between two quantum gates and two convolutional layers. (a) Two parallel two-qubit quantum gates, denoted as $U$ and $B$, are applied to the four least significant qubits. The input state $\left | \varphi \right \rangle _{in} $ is prepared in the same manner as described in Fig. \ref{Fig1}. (b) The matrix transformation of the quantum circuit in (a) can be expressed as gate $B$ acting on the output state of gate $U$ (labeled as $\left | \varphi \right \rangle $ in the figure). The output state of gate $U$ is derived as shown in Fig. \ref{Fig1}. Gate $B$ is represented as a block matrix comprising $2^{m-4}$ blocks along the diagonal, where each block corresponds to the matrix $B \otimes I^{\otimes2}$. Each block processes the pixels of a local region containing four unit regions. Specifically, the first block processes the first four forth-nearest neighbor elements (i.e., the $1$st to $4$th, $5$th to $8$th, $9$th to $12$th and $13$th to $16$th pixels, labeled as $1\xrightarrow{}4, 5\xrightarrow{}8, 9\xrightarrow{}12, 13\xrightarrow{}16$ in the figure), generating the first feature (labeled F1CH1). This operation is equivalent to performing a second convolution based on the first convolution carried out by gate $U$. Note that the parameters in second, third, and fourth row vectors of the matrix are identical to those in the first row vector (namely all being $b_{11}, b_{12}, b_{13}, b_{14}$), so they should not be considered as distinct channels. However, the fifth, ninth, and thirteenth row vectors serve as different kernels, generating features F1CH2, F1CH3, and F1CH4, respectively. Similarly, the second block processes the next set of pixels ($17$th to $20$th, $21$th to $24$th, $25$th to $28$th and $29$th to $32$th) to produce the second feature (F2CH1, F2CH2, F2CH3, and F2CH4).}
    \label{Fig2}
\end{figure*}

This idea is graphically illustrated in Fig. \ref{Fig2}, using two-qubit gates as an example. The first gate, denoted as $U$, performs a convolution operation over the four nearest-neighbor input data points, as depicted in Fig. \ref{Fig1}. Following this, based on Theorem 1, the second gate, labeled $B$, convolves the four forth-nearest-neighbor points of the output amplitudes of the first gate. That is, these two sequential gates create a more abstract representation of the 16 nearest-neighbor input data points, effectively mirroring the functionality of two stacked convolutional layers in classical CNNs.  

Building upon this principle, introducing a third quantum gate after gate $B$ enables the quantum circuit to emulate the behavior of three convolutional convolutional layers, producing a hierarchical feature representation of the 64 nearest-neighbor input data points. This iterative process shows how multiple quantum gates can be leveraged to implement multiple convolution layers, achieving hierarchical feature extraction analogous to that of classical CNNs.

In general, the index positon of a specific output feature in the quantum state amplitude is formulated as follows: 

\textbf{Theorem 2.} Sequentially applying $k$ $n$-qubit quantum gates, from least-significant to most-significant qubits, on an $m$-qubit input state is mathematically equivalent to performing $k$ convolutional layers. The $j$th output feature of channel $c$, denoted as $F_{j,c}$, is encoded in the amplitude of the basis state $\left | i \right \rangle$ of the output state. The index $i$ is given by:
\begin{equation}\label{Eq2}
   \begin{aligned}
     & i=2^n \cdot 2^{n(k-1)} \cdot (j-1) + 2^{n(k-1)} \cdot (c-1),\\
     & \text{where } j=1,2,...,\frac{2^{m-n}}{2^{n(k-1)}}, \\
     & \text{and } c=1,2,...,2^n, k=1,2,...,\left\lfloor \frac{m}{n} \right\rfloor.
   \end{aligned}
\end{equation}

\subsubsection*{Nonlinearity}
\noindent The $k$ parallel quantum gates in Theorem 2 generally perform unitary transformations without introducing any nonlinearity between the convolutional layers. However, in classical CNNs, nonlinear activation functions between convolutional layers are essential for extracting deep and hierarchical features. According to Theorem 1, each quantum gate generates multiple channels, and as stated in Theorem 2, these channels are encoded in the amplitudes of specific basis states. To incorporate a form of nonlinearity between layers in the quantum framework, certain partial channels can be selectively chosen and transformed independently.

\begin{figure*}
    \centering
    \includegraphics[width=1.0\linewidth]{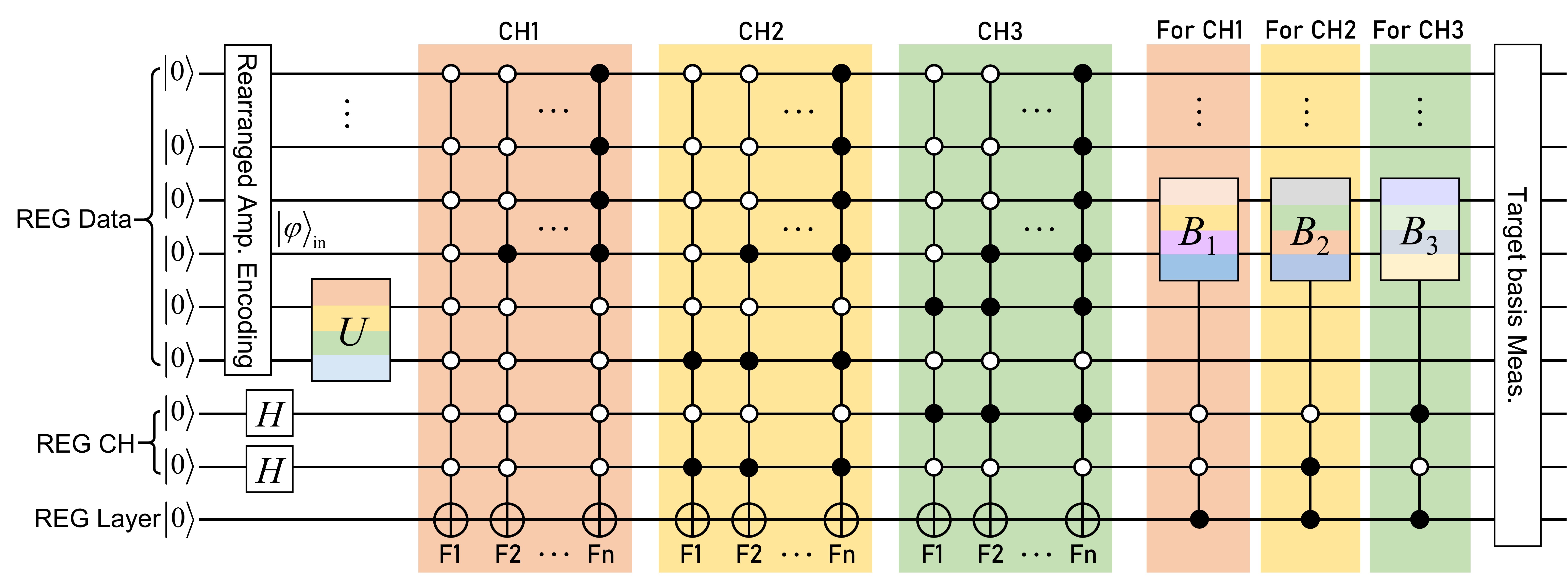}
    \caption{Illustration of the method for introducing nonlinearity between convoultional layers. The data register (labeled REG Data) encodes image pixel values using the method described in Fig. \ref{Fig1}. The gate $U$ is used to implement the first convolutional layer. As depicted in Fig. \ref{Fig1} and formally described in Theorem 2, the features (F1, F2, ..., Fn) of channel 1 (labeled CH1) are encoded in the amplitudes of the basis states $(\left | 0 \right \rangle, \left | 4 \right \rangle, \left | 8 \right \rangle, \left | 12 \right \rangle, ...)$, while the features of channel 2 (CH2) are encoded in $(\left | 1 \right \rangle, \left | 5 \right \rangle, \left | 9 \right \rangle, \left | 13 \right \rangle, ...)$, and so forth. Accordingly, features in each channel can be selectively extracted through controlled operations performed by the data register. The channel register (REG CH) is employed to distinguish between channels, with states such as $\left | 00 \right \rangle$ representing channel 1, $\left | 01 \right \rangle$ representing channel 2, and so on. The layer register (REG Layer) is used to implement multiple convolutional layers, with each additional layer requiring a dedicated layer register. The channel and layer registers are then combained to enable distinct transformations for different channels, such as applying $B_1$ to channel 1, $B_2$ to channel 2, and so forth. After two convolutional layers, the resulting features are encoded in specific basis states. Finally, computational basis measurements are performed to extract the feature values.}
    \label{Fig3}
\end{figure*}

This selection process is achieved through controlled operations on ancilla qubits. Fig. \ref{Fig3} illustrates a scenario in which two convolutional layers are constructed using two-qubit gate kernels. The channel register (denoted as REG CH) marks the target channels, while the layer register (denoted as REG Layer), in conjunction with the channel register, controls the application of subsequent convolutional layers to the selected channels. Features in the target channels are extracted through controlled operations performed by the data register, with the control mechanism designed based on the conclusions established in Theorem 2.

Finally, based on Theorem 1 and Theorem 2, we present several remarks regarding the QCNN model proposed by Cong et al.\textsuperscript{\cite{ref15}}. In this model, each quantum convolutional layer typically consists of $k$ parallel, uniformly parametrized two-qubits quantum gates, where $k=\left\lfloor \frac{m}{2} \right\rfloor$ and $m$ represents the number of qubits in the circuit. According to Theorem 2, such a quantum convolutional layer actually performs $k$ classical convolutional layers, albeit without introducing nonlinearity between layers. However, the shared parameters among the $k$ gates result in the corresponding $k$ classical convolutional layers employing identical kernels. This parameter-sharing mechanism fundamentally diverges from that of classical CNNs. It fails to capture the structural features of local connectivity and sparse connections, and significantly reduces the number of independent channels. These limitations hinder the model's capability for hierarchical feature extraction.

Furthermore, the pooling layer in this QCNN is implemented using controlled gates between adjacent qubits combined with mid-circuit measurements. However, as indicated by Theorem 2, the features of each output channel $F_{j,c}$ are encoded in the amplitude of specific basis states $\left | i \right \rangle$. The controlled gates in the pooling layer must be carefully designed to extract and transform the feactures for each specific channel. Therefore, this current QCNN design obscures the natural convolutional processes inherent in QNNs, significantly limiting its effectiveness in processing grid-like structured data, such as images.

\subsection*{Numerical Experiments}
\noindent To demonstrate the inherent convolution properties of QNNs, we propose novel QCNN architectures that explicitly leverage these properties and validate their effectiveness in image data processing, particularly in the task of multiclass image classification. For this purpose, the widely used MINIST dataset is employed for training and testing. In the experiments, the QCNN architectures comprise two convolutional layers. To evaluate the impact of channel count on model performance, varying numbers of channels\textemdash specifically 1, 4, and 8 channels\textemdash are incorporated between the two layers. Based on preliminary experiments, the convolution kernel size is set to $4 \times 4$, corresponding to the use of four-qubit quantum gates in the quantum circuit. Detailed circuit implementations of the models are described in the Method section. 

All numerical experiments are conducted using the QPanda and pyVQNet 2.0.7 software frameworks\textsuperscript{\cite{ref35}} on a 13th Gen $\rm Intel^\circledR$ $\rm Core^{\text{\tiny TM}}$ i9-13900KF CPU. Each QCNN model is trained for 10 epochs, followed by performance evaluation on the test set. The training process employs the Adam optimizer, with cross-entropy as the loss function, and a learning rate of 0.1.

The classification accuracies for 2-class, 4-class, and 8-class classification tasks, achieved using QCNN models with 1, 4, and 8 channels between layers, are presented in Tab. \ref{Tab1}. Before delving into a detailed discussion of these results, we first compare our results with those of existing QNN models, as summarized in Tab. \ref{Tab2}. Note that mainly the results of pure QNN models are involved in Tab. \ref{Tab2}, while the hybrid quantum-classical models using classical parameters are not involved.

As shown in Tab. \ref{Tab2}, unlike other models, our method performs end-to-end learning without relying on classical feature extraction (FE) techniques, such as principal component analysis. This actually places greater demands on the model's intrisic feature extraction capabilities. Despite this, our model achieves high multiclass classification accuracy while utilizing significantly fewer parameters. These results suggest that our QNN model effectively leverages key mechanisms of classical CNNs, including local recptive fields as well as distributed and hierarchical feature representations.

\begin{table}[!htbp]
    \centering
    \renewcommand\arraystretch{1.5}
    \caption{Classification accuracies for binary, four-class, and eight-class image classification tasks using the MINIST dataset. Three QCNN architectures, each employing 1, 4, and 8 channels between the convolutional layers, are tested.}
    \label{Tab1}
    
    \begin{tabular}{p{2cm}<\centering p{1.5cm}<\centering p{1.5cm}<\centering p{1.5cm}<\centering}
         \Xhline{1.2pt}
         \multirow{2}{*}{Class count}& \multicolumn{3}{c}{Channel count} \\
         \cline{2-4}
          & 1 & 4 & 8 \\
         \hline
         2 & 96\% & 92\% & 86\% \\
         4 & 65\% & 73\% & 76\% \\
         8 & 53\% & 55\% & 58\% \\
         \Xhline{1.2pt}
    \end{tabular}
    
\end{table}

\begin{table*}[!htbp]
    \centering
    \renewcommand\arraystretch{1.5}
     \caption{Comparision of classification accuracies for multiclass image classification between our model and other existing QNNs. The FE column indicates whether classical feature extraction techniques are empolyed. The number of parameters used in each model is also listed.}
    \label{Tab2}
    
    \begin{tabular}{ccccccccccccc}
         \Xhline{1.2pt}
         & \multirow{2}{*}{FE} & \multicolumn{2}{c}{2-class} & & \multicolumn{2}{c}{4-class} & & \multicolumn{2}{c}{8-class} & & \multicolumn{2}{c}{9-class} \\
         \cline{3-4} \cline{6-7} \cline{9-10} \cline{12-13}
         & & Params & Acc. & & Params & Acc. & & Params & Acc. & & Params & Acc. \\
         \hline
         Hur\textsuperscript{\cite{ref29}} & Yes & 12-51 & $96\pm4\% $ & & - & - & & - & - & & - & - \\
         Mahmud\textsuperscript{\cite{ref32}} & Yes & 50 & 98\%-99\% & & - & - & & - & - & & - & - \\
         Smaldone\textsuperscript{\cite{ref31}} & Yes & $>$5200 & 85\%-93.5\% & & $>$5200 & 30\%-40\% & & $>$5200 & 28.2\%-37\% \\
         Du\textsuperscript{\cite{ref12}} & Yes & 54 & 100\% & & - & - & & - & - & & 1500 & 50\% \\
         This study & No & 40 & 96\% & & 40 & 76\% & & 40 & 58\% & & 40 & 55\% \\
         \Xhline{1.2pt}
    \end{tabular}
   
\end{table*}

As shown in Tab. \ref{Tab1}, the classification accuracy decreases as the number of categories increases, while increasing the number of channels can effectively improve multiclass classification accuracy. This indicates that a greater number of channels improves the model's feature extraction capability, which is crucial for addressing multiclass classification tasks. To further substantiate this observation, for the model with 1 channel, we extract features from eight channels during the measurement phase and process them separately. This improves classification accuracy from $65 \%$ to $85 \%$.

On the other hand, in binary classification tasks, an increase in the number of channels leads to a decline in classification accuracy. This counterintuitive result arises because adding more channels increases the number of circuit qubits. Due to the normalization property of quantum states, the amplitudes of specific computational bases diminishes, which in turn exacerbates gradient errors and slows parameter updates, ultimately degrading the model’s performance. These findings highlight the necessity of balancing the model's feature extraction capability and optimization efficiency when determining the optimal number of channels employed in QNNs.

\begin{figure}[htbp]
    \centering
    \includegraphics[width=0.98\linewidth]{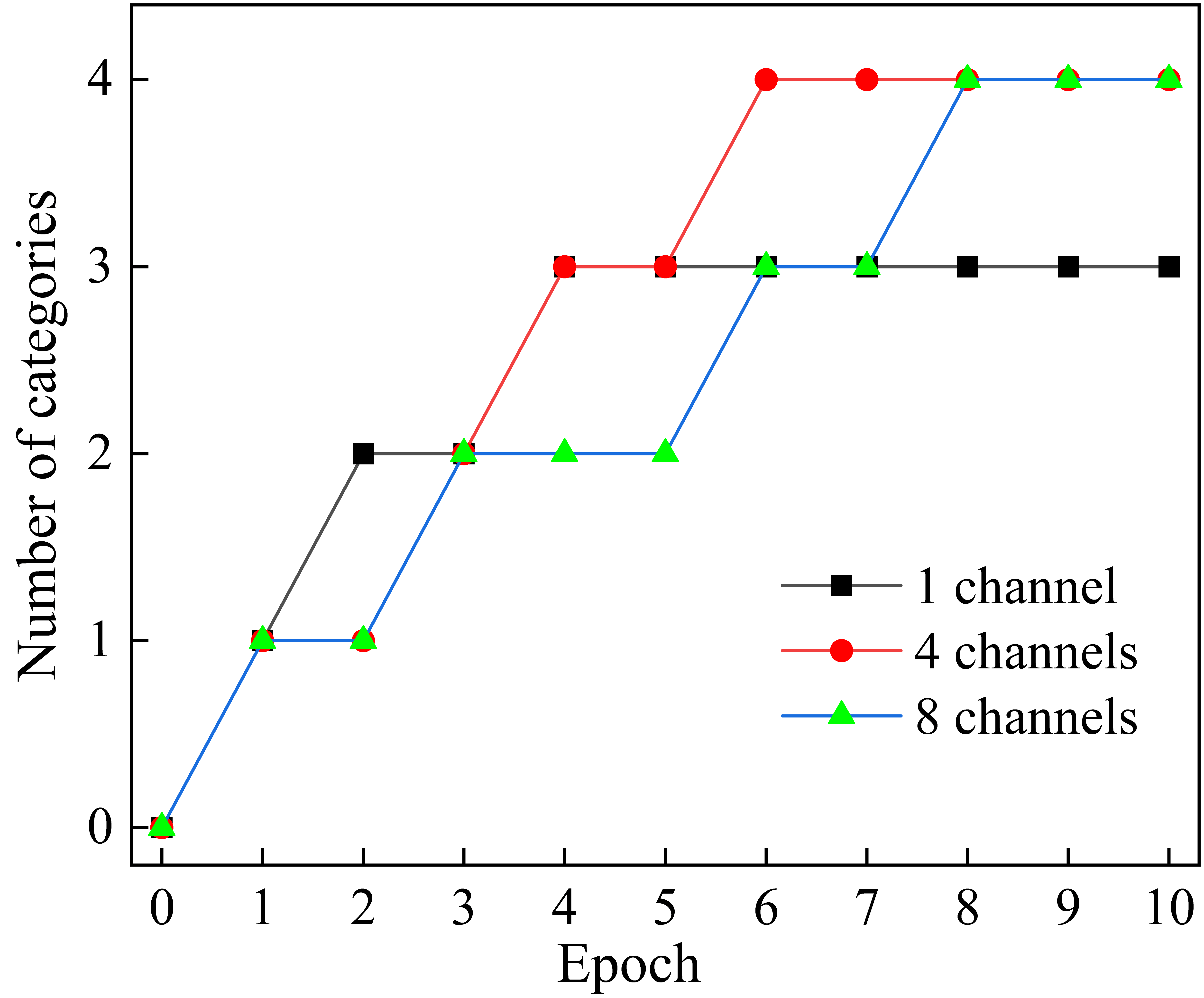}
    \caption {The curve of the number of categories learned by the QNN model with the increase of the number of epochs in the 4-class classification tasks.}
    \label{Fig4}
\end{figure}

\begin{figure}[htbp]
    \centering
    \includegraphics[width=0.90\linewidth]{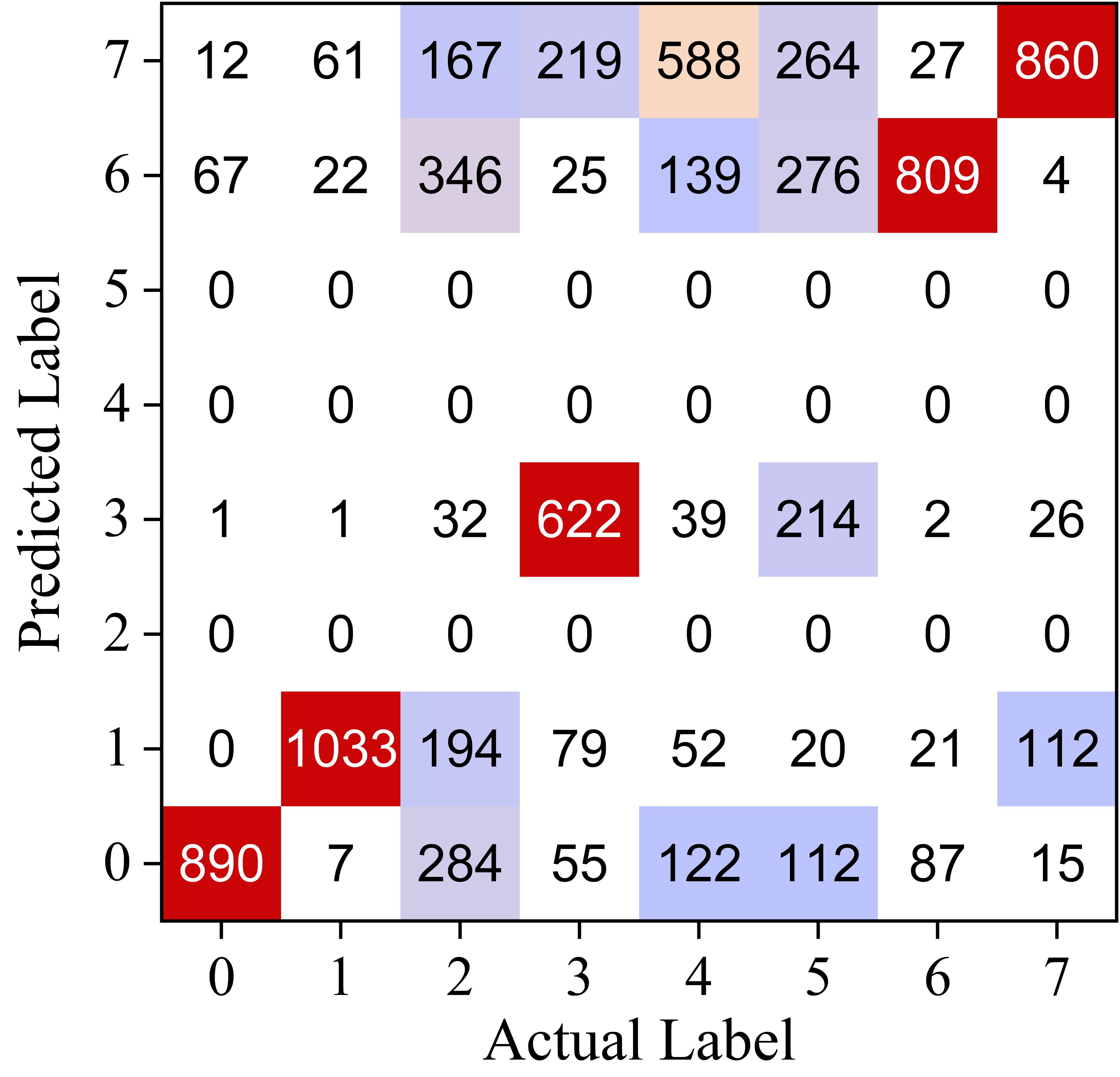}
    \caption{The confusion matrix of the 8-class classification results using 4 channels.}
    \label{Fig5}
\end{figure}

We also find that during the training process, our model seem to learn image categories sequentially. As shown in Fig. \ref{Fig4}, the number of categories learned increases progressively with the number of epochs. The multi-channel model learns new categories more slowly compared to the single-channel model. This is due to the reasons mentioned earlier: the larger number of qubits in the multi-channel model results in smaller amplitudes of the target computational basis states and greater parameter gradient errors. However, the multi-channel model ultimately learns more categories because of its superior feature extraction capabilities. 

\begin{figure*}
    \centering
    \includegraphics[width=1.0\linewidth]{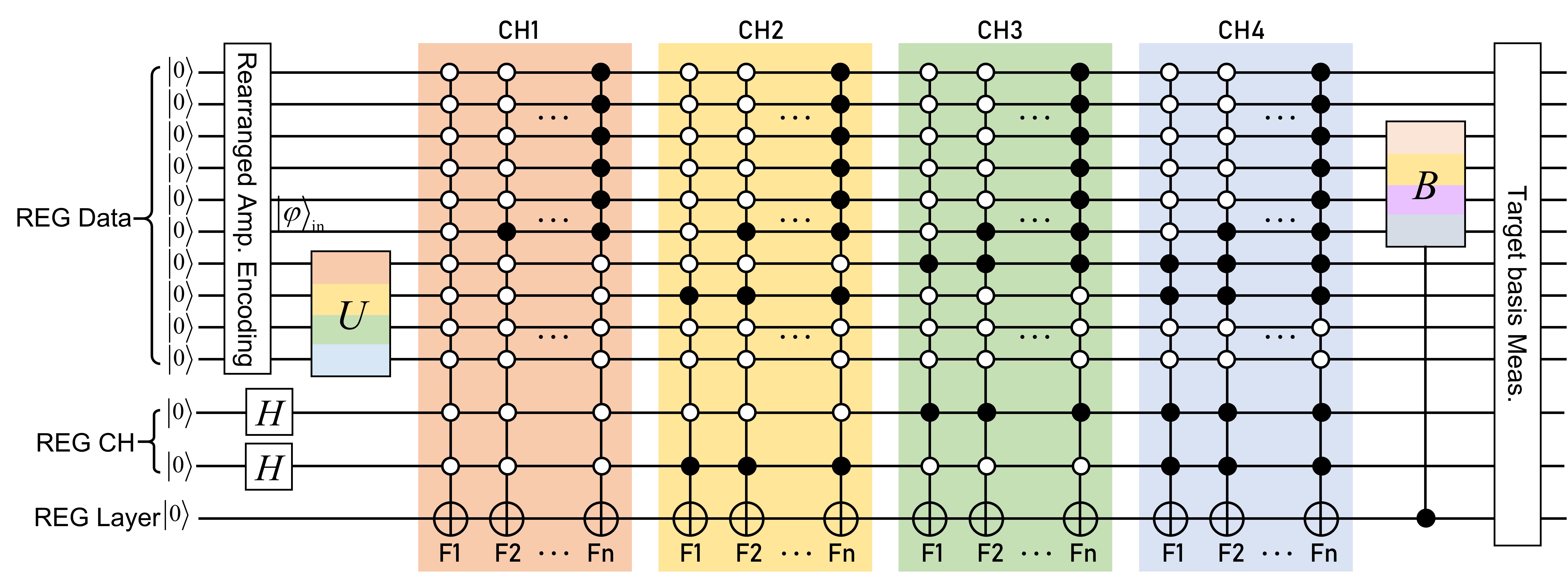}
    \caption{Quantum circuit of the QNN model with 4 channels, which is used in the numerical experiments.}
    \label{Fig6}
\end{figure*}

Moreover, the order in which categories are learned follows a specific pattern. Categories at the two ends of the range are generally learned earlier, while those in the middle are more challenging to learn. As shown in Fig. \ref{Fig5}, after 10 epochs, categories 0, 1, 6, and 7 are learned first, while categories 2, 4, and 5 have not yet been learned. Note that these categories are not related to the actual labels of the images but correspond to encoded numerical categories. This phenomenon should arise from the probability distribution of bit strings generated by the quantum circuit outputs.

\begin{figure}[h]
    \centering
    \includegraphics[width=0.91\linewidth]{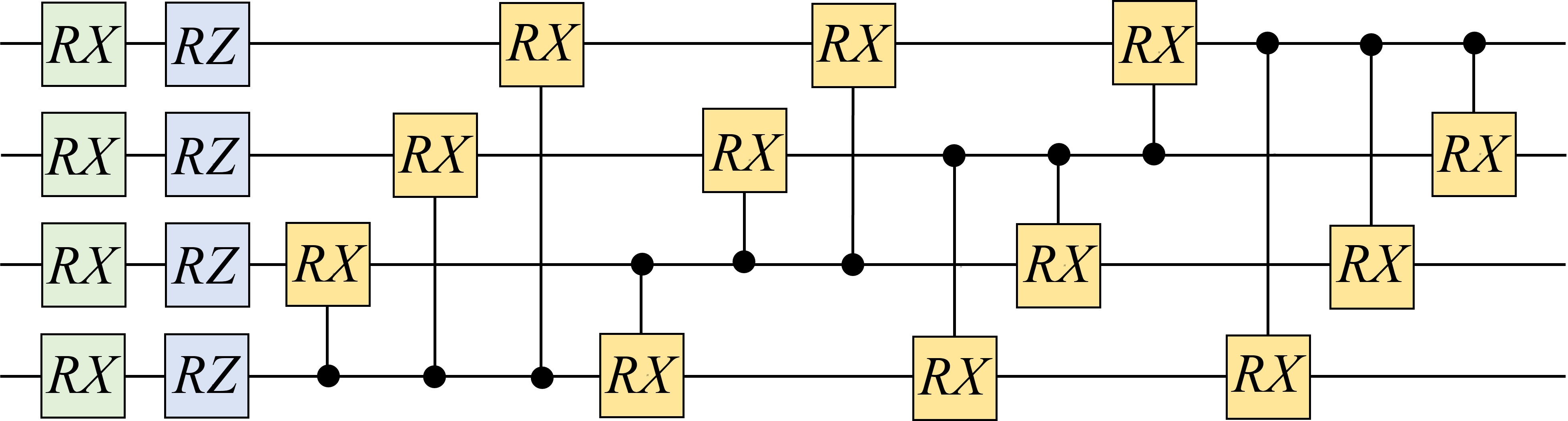}
    \caption {Parameterized quantum circuit for implementing the $U$ and $B$ operations used in the circuit of Fig. \ref{Fig6}.}
    \label{Fig7}
\end{figure}

\subsection*{Methods}
\subsubsection*{Data encoding}
\noindent In our numerical experiments, the learning dataset is the MINIST dataset. To simplify date encoding, the image size is expanded from $28 \times 28$ to $32 \times 32$ by zero-padding along the image edges. The resulting 1024 pixels are normalized and encoded into a quantum state using the amplitude encoding method, as illustrated in Fig. \ref{Fig1}. Note that in the experiments, the size of the convolutional kernel is set to $4 \times 4$, which means each $4 \times 4$ local region of the image constitutes a rearranged unit.

\subsubsection*{Quantum circuits} 
\noindent In the experiments, we evaluate three models with 1, 4, and 8 channels between the two convolutional layers. The architectures of the quantum circuits corresponding to these models are based on the design shown in Fig. \ref{Fig3}. Specifically, the data register consists of 10 qubits to encode the 1024 pixels of each image, while the layer register contains 1 qubit to implement the second convolutional layers. The channel register is configured with 1, 2, and 3 qubits for the models with 1, 4, and 8 channels, respectively. For example, Fig. \ref{Fig6} shows the quantum circuits with 4 channels. For the sake of simplicity, only one transformation $B$ is used in the second convolutional layer. The $U$ and $B$ operations are implemented using four-qubits PQCs. Various PQC designs have been proposed in the literature\textsuperscript{\cite{ref37}}, and the specific PQC employed in our experiments is shown in Fig. \ref{Fig7}. 


\subsubsection*{Post-processing}
\noindent According to Theorem 2, after the two convolutional layers, the model produces 16 channels, each containing 4 features. The features are encoded into the amplitudes of the basis states of the data register. Specifically, the first feature of the 16 channels is encoded in the basis states $(\left | 0 \right \rangle, \left | 16 \right \rangle, \left | 32 \right \rangle, ..., \left | 240 \right \rangle)$, and the second feature in $(\left | 256 \right \rangle, \left | 272 \right \rangle, \left | 288 \right \rangle, ..., \left | 496 \right \rangle)$, and so on. These feature values are extracted through quantum measurement, and the features from different channels are aggregated to produce the final feature representation. A classical Softmax layer is then applied to perform the classification task.

\bibliographystyle{ieeetr}
\bibliography{ref}

\begin{thebibliography}{10}

\bibitem{ref1}
M.~Cerezo, A.~Arrasmith, R.~Babbush, S.~C. Benjamin, S.~Endo, K.~Fujii, J.~R. McClean, K.~Mitarai, X.~Yuan, L.~Cincio, {\em et~al.}, ``Variational quantum algorithms,'' {\em Nature Reviews Physics}, vol.~3, no.~9, pp.~625--644, 2021.

\bibitem{ref2}
K.~Bharti, A.~Cervera-Lierta, T.~H. Kyaw, T.~Haug, S.~Alperin-Lea, A.~Anand, M.~Degroote, H.~Heimonen, J.~S. Kottmann, T.~Menke, {\em et~al.}, ``Noisy intermediate-scale quantum algorithms,'' {\em Reviews of Modern Physics}, vol.~94, no.~1, p.~015004, 2022.

\bibitem{ref3}
G.~A. Quantum, Collaborators*†, F.~Arute, K.~Arya, R.~Babbush, D.~Bacon, J.~C. Bardin, R.~Barends, S.~Boixo, M.~Broughton, B.~B. Buckley, {\em et~al.}, ``Hartree-fock on a superconducting qubit quantum computer,'' {\em Science}, vol.~369, no.~6507, pp.~1084--1089, 2020.

\bibitem{ref4}
K.~H. Wan, O.~Dahlsten, H.~Kristj{\'a}nsson, R.~Gardner, and M.~Kim, ``Quantum generalisation of feedforward neural networks,'' {\em npj Quantum information}, vol.~3, no.~1, p.~36, 2017.

\bibitem{ref5}
E.~Farhi and H.~Neven, ``Classification with quantum neural networks on near term processors,'' {\em arXiv: Quantum Physics}, 2018.

\bibitem{ref5-1}
M.~Schuld and N.~Killoran, ``Quantum machine learning in feature hilbert spaces,'' {\em Physical review letters}, vol.~122, no.~4, p.~040504, 2019.

\bibitem{ref6}
M.~Schuld, A.~Bocharov, K.~M. Svore, and N.~Wiebe, ``Circuit-centric quantum classifiers,'' {\em Physical Review A}, vol.~101, no.~3, p.~032308, 2020.

\bibitem{ref7}
K.~Beer, D.~Bondarenko, T.~Farrelly, T.~J. Osborne, R.~Salzmann, D.~Scheiermann, and R.~Wolf, ``Training deep quantum neural networks,'' {\em Nature communications}, vol.~11, no.~1, p.~808, 2020.

\bibitem{ref8}
X.~Pan, Z.~Lu, W.~Wang, Z.~Hua, Y.~Xu, W.~Li, W.~Cai, X.~Li, H.~Wang, Y.-P. Song, {\em et~al.}, ``Deep quantum neural networks on a superconducting processor,'' {\em Nature Communications}, vol.~14, no.~1, p.~4006, 2023.

\bibitem{ref9}
A.~Abbas, D.~Sutter, C.~Zoufal, A.~Lucchi, A.~Figalli, and S.~Woerner, ``The power of quantum neural networks,'' {\em Nature Computational Science}, vol.~1, no.~6, pp.~403--409, 2021.

\bibitem{ref10}
Y.~Du, M.-H. Hsieh, T.~Liu, S.~You, and D.~Tao, ``Learnability of quantum neural networks,'' {\em PRX quantum}, vol.~2, no.~4, p.~040337, 2021.

\bibitem{ref11}
M.~C. Caro, H.-Y. Huang, M.~Cerezo, K.~Sharma, A.~Sornborger, L.~Cincio, and P.~J. Coles, ``Generalization in quantum machine learning from few training data,'' {\em Nature communications}, vol.~13, no.~1, p.~4919, 2022.

\bibitem{ref12}
Y.~Du, Y.~Yang, D.~Tao, and M.-H. Hsieh, ``Problem-dependent power of quantum neural networks on multiclass classification,'' {\em Physical Review Letters}, vol.~131, no.~14, p.~140601, 2023.

\bibitem{ref12-1}
E.~Gil-Fuster, J.~Eisert, and C.~Bravo-Prieto, ``Understanding quantum machine learning also requires rethinking generalization,'' {\em Nature Communications}, vol.~15, no.~1, p.~2277, 2024.

\bibitem{ref13}
V.~Havl{\'\i}{\v{c}}ek, A.~D. C{\'o}rcoles, K.~Temme, A.~W. Harrow, A.~Kandala, J.~M. Chow, and J.~M. Gambetta, ``Supervised learning with quantum-enhanced feature spaces,'' {\em Nature}, vol.~567, no.~7747, pp.~209--212, 2019.

\bibitem{ref14}
Y.~LeCun, Y.~Bengio, and G.~Hinton, ``Deep learning,'' {\em nature}, vol.~521, no.~7553, pp.~436--444, 2015.

\bibitem{ref15}
I.~Cong, S.~Choi, and M.~D. Lukin, ``Quantum convolutional neural networks,'' {\em Nature Physics}, vol.~15, no.~12, pp.~1273--1278, 2019.

\bibitem{ref16}
M.~Henderson, S.~Shakya, S.~Pradhan, and T.~Cook, ``Quanvolutional neural networks: powering image recognition with quantum circuits,'' {\em Quantum Machine Intelligence}, vol.~2, no.~1, p.~2, 2020.

\bibitem{ref17}
J.~Herrmann, S.~M. Llima, A.~Remm, P.~Zapletal, N.~A. McMahon, C.~Scarato, F.~Swiadek, C.~K. Andersen, C.~Hellings, S.~Krinner, {\em et~al.}, ``Realizing quantum convolutional neural networks on a superconducting quantum processor to recognize quantum phases,'' {\em Nature communications}, vol.~13, no.~1, p.~4144, 2022.

\bibitem{ref17-1}
Y.~Li, R.-G. Zhou, R.~Xu, J.~Luo, and W.~Hu, ``A quantum deep convolutional neural network for image recognition,'' {\em Quantum Science and Technology}, vol.~5, no.~4, p.~044003, 2020.

\bibitem{ref17-2}
I.~Kerenidis, J.~Landman, and A.~Prakash, ``Quantum algorithms for deep convolutional neural networks,'' {\em arXiv preprint arXiv:1911.01117}, 2019.

\bibitem{ref17-3}
S.~Wei, Y.~Chen, Z.~Zhou, and G.~Long, ``A quantum convolutional neural network on nisq devices,'' {\em AAPPS bulletin}, vol.~32, pp.~1--11, 2022.

\bibitem{ref18}
J.~Bausch, ``Recurrent quantum neural networks,'' {\em Advances in neural information processing systems}, vol.~33, pp.~1368--1379, 2020.

\bibitem{ref19}
Y.~Takaki, K.~Mitarai, M.~Negoro, K.~Fujii, and M.~Kitagawa, ``Learning temporal data with a variational quantum recurrent neural network,'' {\em Physical Review A}, vol.~103, no.~5, p.~052414, 2021.

\bibitem{ref20}
Y.~Li, Z.~Wang, R.~Han, S.~Shi, J.~Li, R.~Shang, H.~Zheng, G.~Zhong, and Y.~Gu, ``Quantum recurrent neural networks for sequential learning,'' {\em Neural Networks}, vol.~166, pp.~148--161, 2023.

\bibitem{ref21}
Y.~Li, Z.~Wang, R.~Xing, C.~Shao, S.~Shi, J.~Li, G.~Zhong, and Y.~Gu, ``Quantum gated recurrent neural networks,'' {\em IEEE Transactions on Pattern Analysis and Machine Intelligence}, vol.~47, no.~4, 2025.

\bibitem{ref22}
S.~Lloyd and C.~Weedbrook, ``Quantum generative adversarial learning,'' {\em Physical review letters}, vol.~121, no.~4, p.~040502, 2018.

\bibitem{ref23}
P.-L. Dallaire-Demers and N.~Killoran, ``Quantum generative adversarial networks,'' {\em Physical Review A}, vol.~98, no.~1, p.~012324, 2018.

\bibitem{ref24}
M.~S. Rudolph, N.~B. Toussaint, A.~Katabarwa, S.~Johri, B.~Peropadre, and A.~Perdomo-Ortiz, ``Generation of high-resolution handwritten digits with an ion-trap quantum computer,'' {\em Physical Review X}, vol.~12, no.~3, p.~031010, 2022.

\bibitem{ref25}
Y.~Bengio, A.~Courville, and P.~Vincent, ``Representation learning: A review and new perspectives,'' {\em IEEE transactions on pattern analysis and machine intelligence}, vol.~35, no.~8, pp.~1798--1828, 2013.

\bibitem{ref25-1}
J.~R. McClean, S.~Boixo, V.~N. Smelyanskiy, R.~Babbush, and H.~Neven, ``Barren plateaus in quantum neural network training landscapes,'' {\em Nature communications}, vol.~9, no.~1, p.~4812, 2018.

\bibitem{ref25-2}
Z.~Holmes, K.~Sharma, M.~Cerezo, and P.~J. Coles, ``Connecting ansatz expressibility to gradient magnitudes and barren plateaus,'' {\em PRX Quantum}, vol.~3, no.~1, p.~010313, 2022.

\bibitem{ref25-3}
M.~Cerezo, A.~Sone, T.~Volkoff, L.~Cincio, and P.~J. Coles, ``Cost function dependent barren plateaus in shallow parametrized quantum circuits,'' {\em Nature communications}, vol.~12, no.~1, p.~1791, 2021.

\bibitem{ref26}
A.~Pesah, M.~Cerezo, S.~Wang, T.~Volkoff, A.~T. Sornborger, and P.~J. Coles, ``Absence of barren plateaus in quantum convolutional neural networks,'' {\em Physical Review X}, vol.~11, no.~4, p.~041011, 2021.

\bibitem{ref27}
K.~Sharma, M.~Cerezo, L.~Cincio, and P.~J. Coles, ``Trainability of dissipative perceptron-based quantum neural networks,'' {\em Physical Review Letters}, vol.~128, no.~18, p.~180505, 2022.

\bibitem{ref33}
J.~Liu, K.~H. Lim, K.~L. Wood, W.~Huang, C.~Guo, and H.-L. Huang, ``Hybrid quantum-classical convolutional neural networks,'' {\em Science China Physics, Mechanics \& Astronomy}, vol.~64, no.~9, p.~290311, 2021.

\bibitem{ref33-1}
E.~H. Houssein, Z.~Abohashima, M.~Elhoseny, and W.~M. Mohamed, ``Hybrid quantum-classical convolutional neural network model for covid-19 prediction using chest x-ray images,'' {\em Journal of Computational Design and Engineering}, vol.~9, no.~2, pp.~343--363, 2022.

\bibitem{ref34}
Y.-y. Feng, Y.~Li, J.~Li, J.~Zhou, and J.-j. Shi, ``Variational shadow quantum circuits assisted quantum convolutional neural network,'' {\em Advanced Quantum Technologies}, p.~2400510, 2025.

\bibitem{ref29-1}
A.~Simen, C.~Flores-Garrigos, N.~N. Hegade, I.~Montalban, Y.~Vives-Gilabert, E.~Michon, Q.~Zhang, E.~Solano, and J.~D. Mart{\'\i}n-Guerrero, ``Digital-analog quantum convolutional neural networks for image classification,'' {\em Physical Review Research}, vol.~6, no.~4, p.~L042060, 2024.

\bibitem{ref28}
Y.-J. Liu, A.~Smith, M.~Knap, and F.~Pollmann, ``Model-independent learning of quantum phases of matter with quantum convolutional neural networks,'' {\em Physical Review Letters}, vol.~130, no.~22, p.~220603, 2023.

\bibitem{ref28-1}
C.~Umeano, A.~E. Paine, V.~E. Elfving, and O.~Kyriienko, ``What can we learn from quantum convolutional neural networks?,'' {\em Advanced Quantum Technologies}, p.~2400325, 2023.

\bibitem{ref29}
T.~Hur, L.~Kim, and D.~K. Park, ``Quantum convolutional neural network for classical data classification,'' {\em Quantum Machine Intelligence}, vol.~4, no.~1, p.~3, 2022.

\bibitem{ref30}
S.~Shi, Z.~Wang, J.~Li, Y.~Li, R.~Shang, G.~Zhong, and Y.~Gu, ``Quantum convolutional neural networks for multiclass image classification,'' {\em Quantum Information Processing}, vol.~23, no.~5, pp.~1--16, 2024.

\bibitem{ref31}
A.~M. Smaldone, G.~W. Kyro, and V.~S. Batista, ``Quantum convolutional neural networks for multi-channel supervised learning,'' {\em Quantum Machine Intelligence}, vol.~5, no.~2, p.~41, 2023.

\bibitem{ref32}
J.~Mahmud, R.~Mashtura, S.~A. Fattah, and M.~Saquib, ``Quantum convolutional neural networks with interaction layers for classification of classical data,'' {\em Quantum Machine Intelligence}, vol.~6, no.~1, p.~11, 2024.

\bibitem{ref34-1}
F.~Yu, ``Multi-scale context aggregation by dilated convolutions,'' {\em arXiv preprint arXiv:1511.07122}, 2015.

\bibitem{ref34-2}
L.-C. Chen, G.~Papandreou, I.~Kokkinos, K.~Murphy, and A.~L. Yuille, ``Deeplab: Semantic image segmentation with deep convolutional nets, atrous convolution, and fully connected crfs,'' {\em IEEE transactions on pattern analysis and machine intelligence}, vol.~40, no.~4, pp.~834--848, 2017.

\bibitem{ref35}
L.~Origin Quantum Computing Technology (Hefei)~Co., ``Vqnet 2.0 tutorial,'' 2022.

\bibitem{ref37}
S.~Sim, P.~D. Johnson, and A.~Aspuru-Guzik, ``Expressibility and entangling capability of parameterized quantum circuits for hybrid quantum-classical algorithms,'' {\em Advanced Quantum Technologies}, vol.~2, no.~12, p.~1900070, 2019.

\end{thebibliography}

\subsection*{Acknowledgements}
\noindent The present work is supported by the Natural Science Foundation of Shandong Province of China (ZR2021ZD19) and the National Natural Science Foundation of China (12005212). We are grateful for the support from Marine Big Data Center of Institute for Advanced Ocean Study of Ocean University of China, and the professional and technical services provided by Yujie Dong. We also thank the technical team from the Origin Quantum Computing Company for their professional services.

\subsection*{Author contributions}
\noindent Z.W., Y.G. and G.Z. conceived and supervised the research. Z.W. and G.Q. developed the methodology and designed the experiments. G.Q. conducted the numerical experiments. Z.W. drafted the manuscript, with all authors providing critical revisions and contributing to its finalization.

\subsection*{Code availability}
\noindent The code used for the numerical experiments carried out for this work is available
upon request from the authors (E-mail: wangzhimin@ouc.edu.cn).

\subsection*{Competing interests}
\noindent The authors declare no competing interests.

\end{document}